# Spatial correlations of charge density wave order across the transition in 2H-$NbSe_2$

Seokjo Hong[1,§], Jaewhan Oh[1,§], Jemin Park[1,§], Woohyun Cho[1], Soyoung Lee[1], Colin Ophus[2], Yeongkwan Kim[1,⁑], Heejun Yang[1,‡], SungBin Lee[1,*] and Yongsoo Yang[1,3,†]

[1]Department of Physics, Korea Advanced Institute of Science and Technology (KAIST), Daejeon 34141, Republic of Korea
[2]Department of Materials Science and Engineering, Stanford University, Stanford, CA 94305, USA
[3]Graduate School of Semiconductor Technology, School of Electrical Engineering, Korea Advanced Institute of Science and Technology (KAIST), Daejeon 34141, Republic of Korea

Charge density waves (CDWs) involve coupled amplitude and phase degrees of freedom, but direct access to local amplitude correlations remains experimentally challenging. Here, we report cryogenic four-dimensional scanning transmission electron microscopy (4D-STEM) measurements of CDW ordering in a 2H-$NbSe_2$ flake of 24 nm thickness, enabled by liquid helium-based cooling. By mapping the spatial distribution of CDW superlattice intensities at nanometer-scale resolution and analyzing their autocorrelations, we extract the temperature-dependent correlation length associated with the local amplitude of the CDW order parameter, independent of global phase coherence. Our results reveal that a finite local CDW amplitude is already established well above the transition temperature. When the system is cooled below the transition temperature down to 20 K, the correlation length extends to nearly 110 nm, and the local CDW amplitude is found to strongly anticorrelate with the local strain field.

Charge density waves (CDWs) are a fundamental broken symmetry state, marked by a spontaneous modulation of electronic density coupled to the lattice [1–12]. Traditional studies have focused on macroscopic signatures, such as gap openings and satellite diffraction peaks [9,13,14]. These signals originate from averaging over large sample areas and represent a combination of both amplitude and phase contributions [14]. In real materials, however, the onset of the CDW amplitude and the establishment of phase coherence often occur at different temperatures, indicating that these two components of the order parameter can evolve independently. In this regard, disentangling amplitude and phase components is crucial for a deeper understanding of CDW transitions, especially in systems dominated by fluctuations, strong correlations, or disorder such as in pseudogap-like or glassy CDW states [15–19].

However, accessing these components experimentally remains a significant challenge. Techniques such as scanning tunneling microscopy (STM) and magnetic force microscopy have been widely used to study CDWs [19–27], but they are inherently surface-sensitive and cannot easily detect ordering beneath the surface [28]. This makes it difficult to compare their results with bulk transport properties. X-ray, neutron and electron diffraction as well as nuclear magnetic resonance have also been actively used to characterize CDWs [13,29–34]. However, most prior studies rely on ensemble-averaged signals, which primarily reveal long-range periodicity and phase coherence, but often lack the spatial resolution necessary to isolate local CDW order and its correlation.

Four-dimensional scanning transmission electron microscopy (4D-STEM) addresses these limitations by enabling real-space mapping of CDW amplitude fluctuations through local diffraction measurements. By recording diffraction patterns at each probe position using a nanometer-scale focused electron beam, 4D-STEM provides pixel-resolved maps of CDW superlattice peak intensities [35], serving as a direct probe of the local amplitude of CDW order.

In this work, we use liquid helium cryogenic 4D-STEM [Fig. 1(a)] to map the spatial distribution of the local CDW amplitude in 2H-$NbSe_2$, whose triple-Q order sets in below $T_{CDW}$ ~30 K [13,29–32]. Unlike materials such as $TaS_2$, where multiple coexisting CDW phases and a first-order transition complicate analysis [36–38], 2H-$NbSe_2$ offers a simpler and cleaner platform for probing intrinsic CDW formation and spatial correlations near the transition. Interestingly, the local CDW amplitude is already established above $T_{CDW}$, evidenced by a nonzero mean and indicative of preformed local domains. Upon cooling, the correlation length grows to approximately 110 nm at 20 K and the local amplitudes are strongly anticorrelated with the local strain field. These findings deviate from conventional mean-field expectations and highlight the importance of independently tracking amplitude correlations in complex electronic systems, particularly where phase coherence alone does not capture the full nature of the ordering.

A single crystal of 2H-$NbSe_2$ was grown using the chemical vapor transport method, and a platelet of approximately 24 nm thickness [measured using electron energy loss spectroscopy (EELS); see Appendix for details] was mechanically exfoliated and transferred onto a holey $SiN_x$ TEM grid [see Fig. 1(b,c); detailed methods are in the Appendix]. The 4D-STEM experiments were conducted using a liquid helium-based cooling holder, with a temperature range of 20 K to 300 K, on a double Cs-corrected TEM (Titan cubed G2, FEI).

At 20 K, the averaged diffraction pattern clearly reveals the formation of 3 × 3 CDW domains, indicated by superlattice peaks at integer multiples of the 1/3 reciprocal lattice vector positions [Fig. 2(a) and Supplemental Movie 1]. These peaks disappear at room temperature [Fig. 2(b)].

It is important to note that not all expected CDW peaks are present in the diffraction pattern at 20 K, likely due to an unavoidable tilt between the normal vector of the $NbSe_2$ van

⁑ Contact author: yeongkwan@kaist.ac.kr
‡ Contact author: h.yang@kaist.ac.kr
* Contact author: sungbin@kaist.ac.kr
† Contact author: yongsoo.yang@kaist.ac.kr
§ These authors contributed equally to this work.

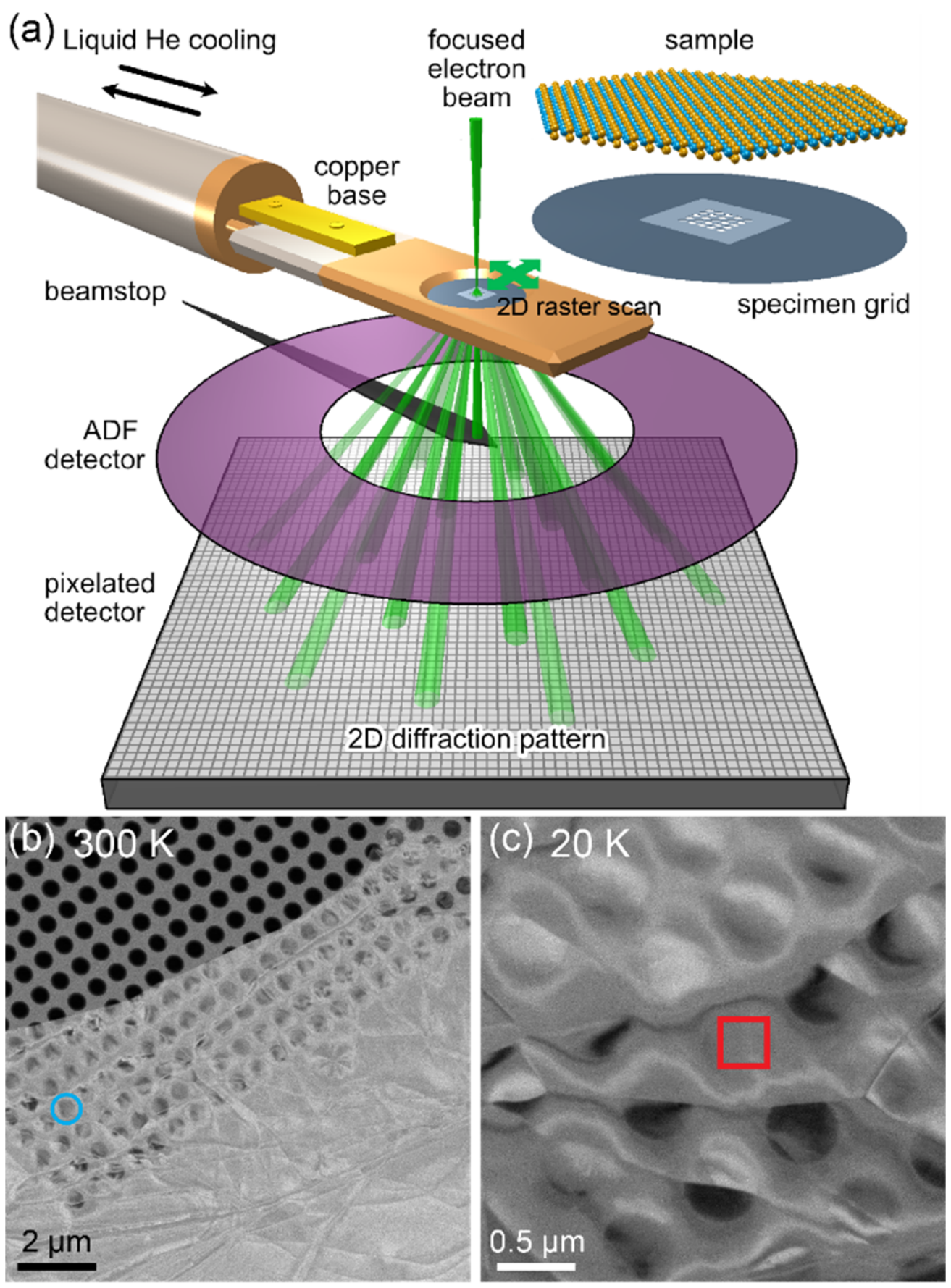

FIG 1. (a) Schematic representation of the low-temperature 4D-STEM experimental setup. (b) Low-magnification annular dark-field STEM (ADF-STEM) image of a holey $SiN_x$ grid, partially covered with exfoliated 2H-$NbSe_2$, with the measured hole highlighted by a blue circle. (c) Higher-magnification ADF-STEM image showing the area analyzed via 4D-STEM, indicated by a red box.

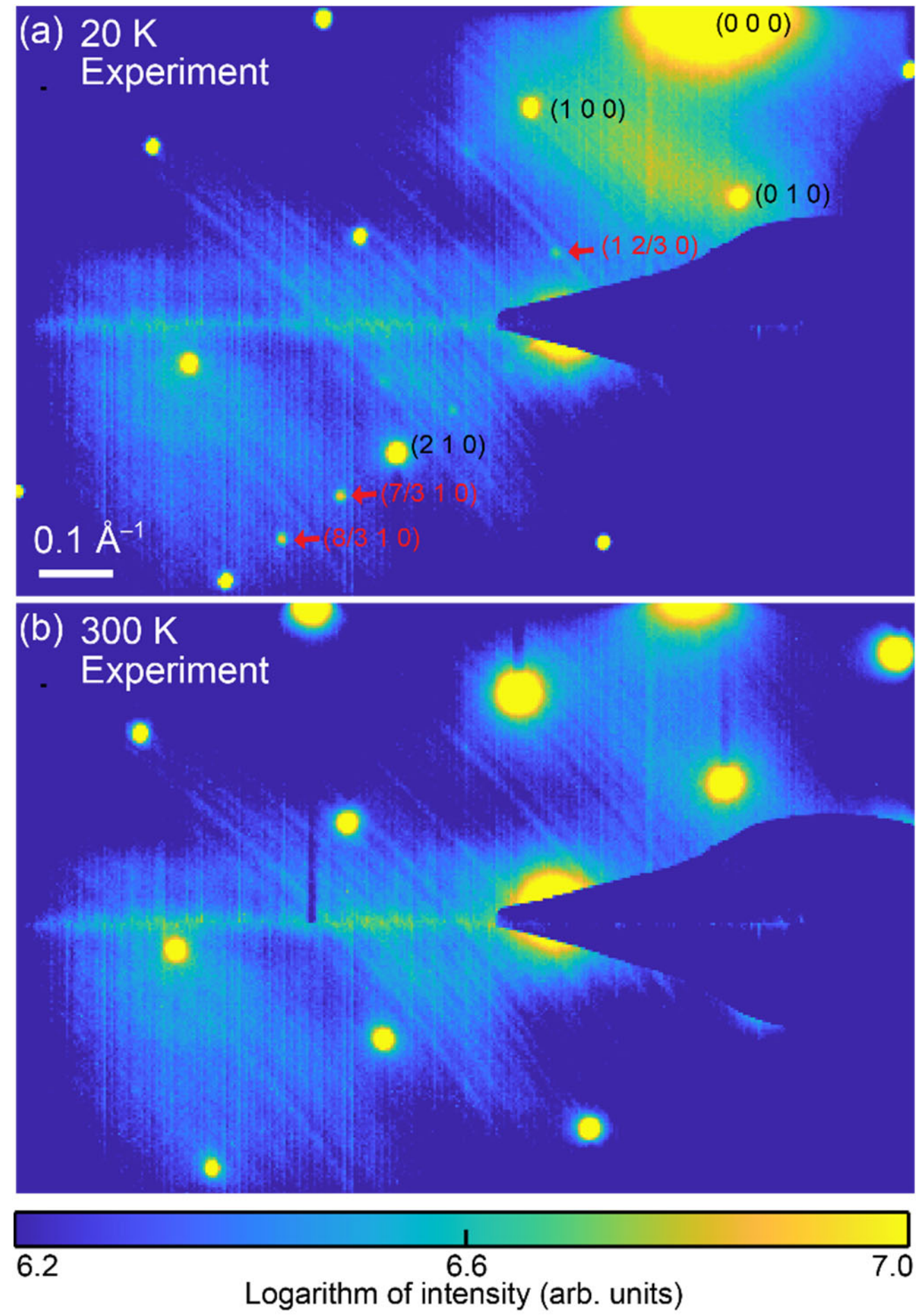

FIG 2. (a) Averaged diffraction pattern obtained from 120 × 120 scanned probe positions (spanning ~ 200 × 200 nm$^2$) in the 4D-STEM dataset acquired at 20 K. Peaks are indexed as (h k l) based on the reciprocal lattice vectors of 2H-$NbSe_2$. (b) Similarly averaged diffraction pattern from the 4D-STEM dataset acquired at 300 K.

der Waals layers and the electron beam direction. To verify this, multislice-based electron scattering simulations (see Appendix for details) were performed, showing that a ~3° tilt results in a diffraction pattern that closely matches the experimentally measured one (see Appendix). This suggests that the absence of certain CDW peaks in the experimental data results from diffraction conditions not being satisfied due to sample tilt.

The CDW Bragg-peak intensity directly measures the order parameter in $NbSe_2$ via the linear coupling of charge modulation and lattice [39]. As the CDW amplitude grows, the superlattice intensity increases. We focus on the particularly prominent peaks at (7/3 1 0) and (8/3 1 0), extracting their intensities across probe positions and temperatures from the diffraction patterns (see Appendix).

Figure 3 presents temperature-resolved maps of local CDW peak intensities from the 4D-STEM data. For both the (7/3 1 0) and (8/3 1 0) peaks, we consistently observe strong but spatially inhomogeneous order at 20 K. With increasing temperature, the intensity weakens and the ordered domains fragment, fading to background by room temperature.

In a clean, weakly coupled mean-field transition, the order parameter is nearly uniform well below $T_{CDW}$. The order parameter vanishes well above the transition, and the correlation length diverges near the critical point. In contrast, we observe substantial spatial variation in CDW peak intensity even well below $T_{CDW}$. Above $T_{CDW}$, finite intensity persists across the scanned region without spatial correlation, appearing as noise-like fluctuations rather than a coherent pattern. This behavior is consistent with previous STM reports of nanoscale CDW domains nucleating near defects and persisting above the transition, as well as X-ray studies that identify coexisting short- and long-range CDW components in $NbSe_2$ [19,22,40]. This continued presence of weak CDW intensity above the transition has been interpreted as a pseudogap-like regime, where local CDW amplitude exists without long-range phase coherence [19].

The 4D-STEM maps of the local CDW amplitude allows us to calculate the spatial autocorrelation $\langle\varphi(\mathbf{r})\varphi(\mathbf{r}_0+\mathbf{r})\rangle$, and to extract correlation lengths as a function of temperature, where $\varphi(\mathbf{r})$ represents the spatial distribution of the CDW order parameter (Fig. 4; see Appendix). At 20 K, well below $T_{CDW}$, the correlation length reaches approximately 110 nm.

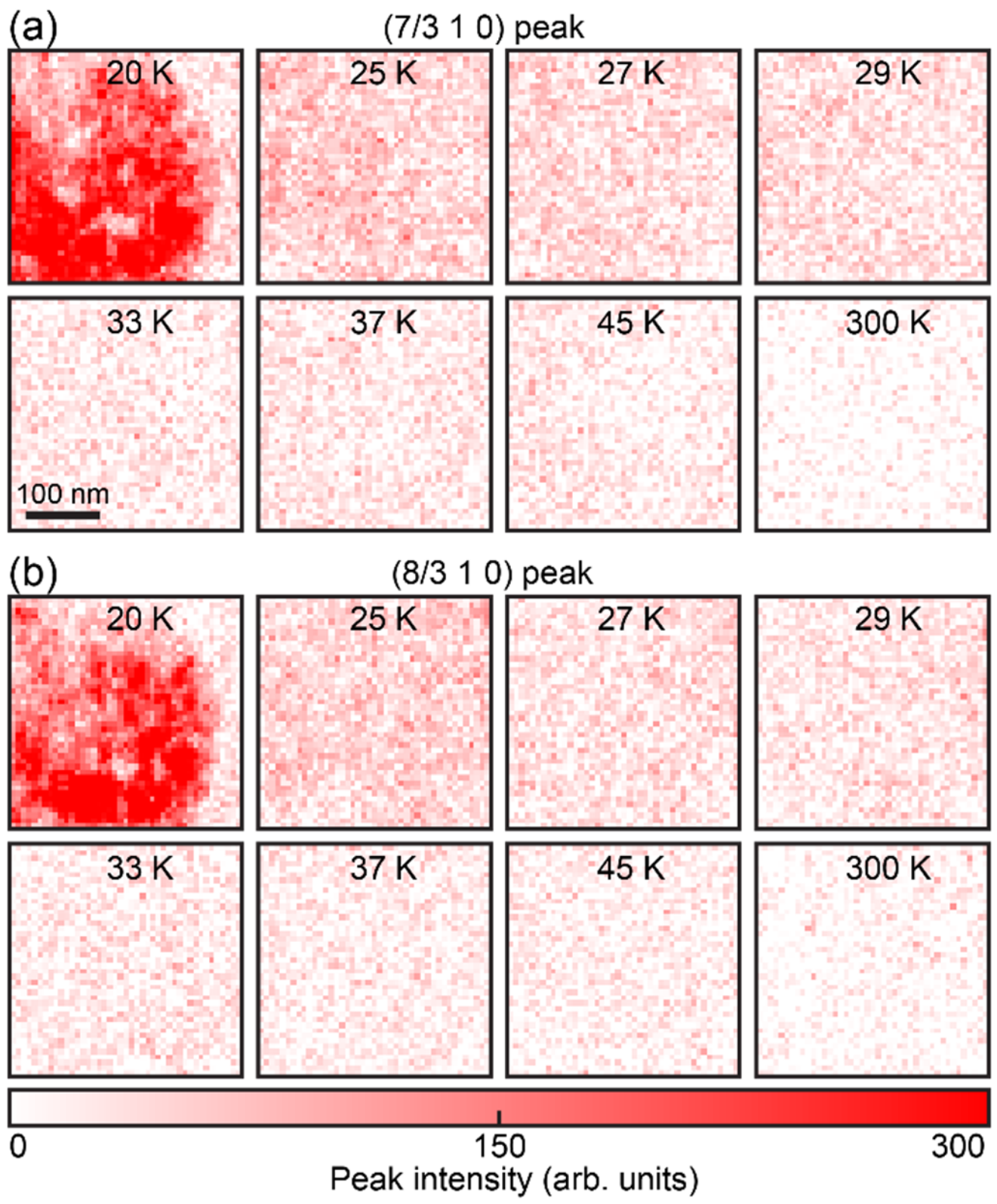

FIG 3. (a) Temperature-dependent spatial mapping of the (7/3 1 0) peak intensity across the scanned region. (b) Corresponding evolution of the (8/3 1 0) peak intensity. Pixels with negative intensity values are displayed as zero intensity for clarity.

Because our map field of view is ~300 nm, and reliable correlation length estimates typically require a system size at least five times larger [41], lengths above ~60 nm are likely underestimated; the 110 nm should therefore be regarded as a lower bound. With increasing temperature, the correlation length collapses to nearly zero above $T_{CDW}$. In contrast, the constant background of the correlation curves, which reflects the mean local CDW amplitude [Fig. 4(a,c)], remains finite and gradually decays, approaching zero only at room temperature. This indicates that amplitude order can persist without global phase coherence, consistent with earlier STM and spectroscopic observations of pseudogap-like CDW states.

It is important to note that each pixel in the CDW map represents the peak intensity from a local electron diffraction pattern measured with a ~12 nm electron probe, providing a reliable measure of the coherent CDW order within the probe area. However, intensity-intensity correlations between probe positions (pixels in Fig. 3) are not directly sensitive to whether they share a common phase. As a result, the extracted correlation length reflects the spatial distribution of local CDW amplitudes rather than global phase coherence. This distinction is essential because it allows us to analyze amplitude correlations even in regimes where long-range phase coherence has not yet developed. Our method thus enables experimental separation of the amplitude and phase sectors of the CDW order parameter.

Another key point is a strong anticorrelation between the local CDW amplitude and the strain field at 20 K, i.e., strain suppresses CDW development. To quantify this effect, we extracted nanoscale strain from the 4D-STEM data (see Appendix). The resulting maps (Fig. 5) show that most of the scanned area has only minor strain (< 0.5 %), yet localized strain hotspots coincide with suppressed CDW intensity, naturally accounting for the finite correlation length observed at 20 K. In addition, we performed numerical simulations with spatially varying strain (details in the Appendix) to corroborate our findings. They reproduce the observed strain-CDW anticorrelation below $T_{CDW}$, show little strain effect above $T_{CDW}$, and capture the emergence of a finite correlation length below the transition, consistent with the experimental behavior. Within a Landau description, $F(\varphi,T) = \alpha\,(T - T_0)|\varphi|^2 + \beta\,|\varphi|^4 + \lambda\,\varepsilon\,|\varphi|^2 + \cdots$, a positive coupling $\lambda$ raises the local mass term due to the strain $\varepsilon$, and thereby suppresses the local CDW amplitude $|\varphi|$, consistent with the measured anticorrelation.

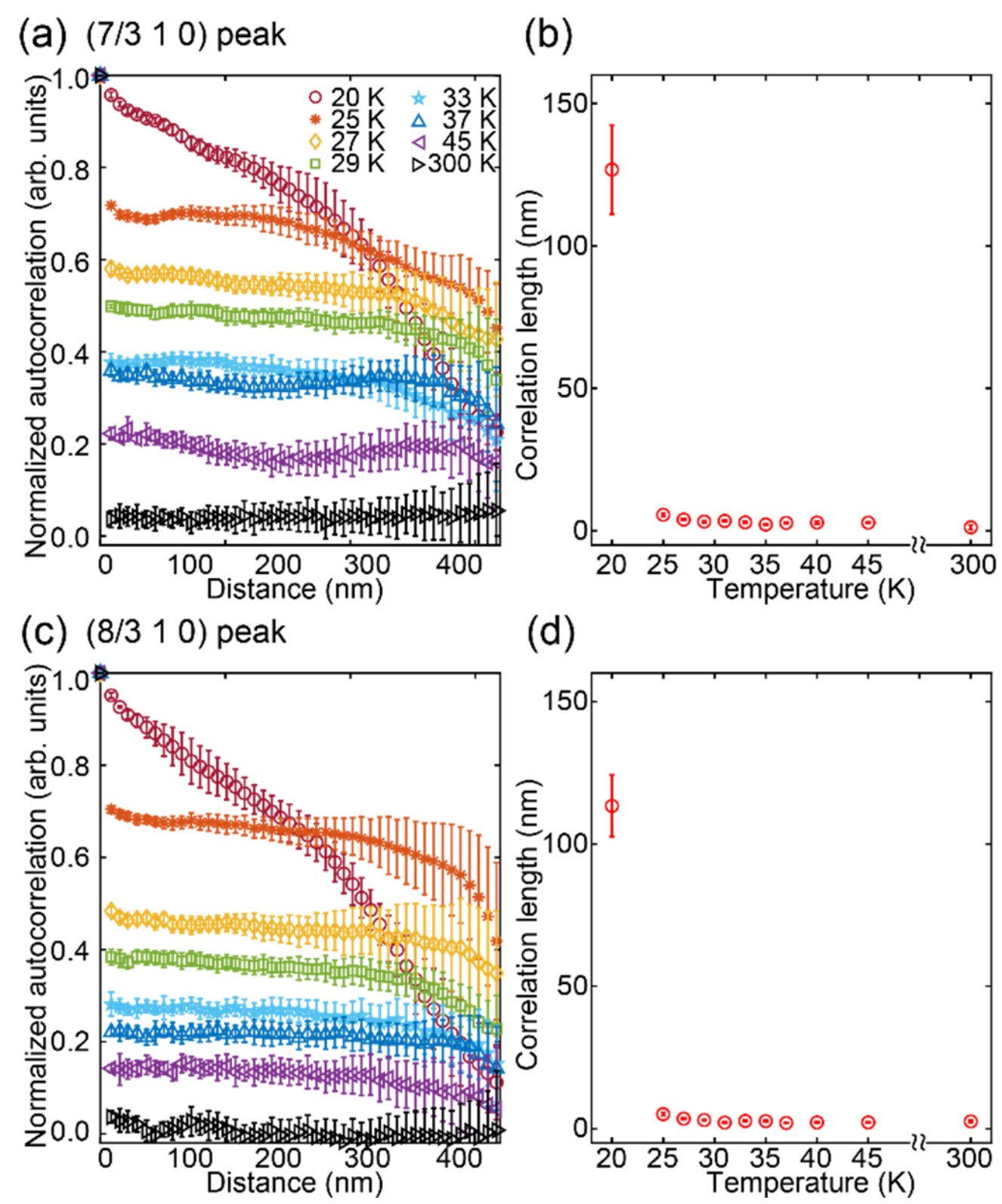

FIG 4. (a) Temperature-dependent radial profiles of the 2D autocorrelation function calculated from the (7/3 1 0) peak intensity map [shown in Fig. 3(a)]. (b) Correlation length as a function of temperature, extracted from the autocorrelation profiles in (a). (c,d) Corresponding results for the (8/3 1 0) peak intensity map, following the same analysis approach as in (a,b).

Unlike surface-sensitive techniques such as STM, our 4D-STEM approach captures diffraction from the entire sample thickness along the electron beam direction. This depth sensitivity enables our measurements to probe spatial correlations relevant to bulk properties and directly comparable to macroscopic transport behavior. A finite value of measured correlation length indicates the role of microscopic inhomogeneity in limiting long-range order. In particular, this length scale likely reflects the typical size of CDW amplitude domains, which can influence transport by locally gapping parts of the Fermi surface. Therefore, the restricted spatial extent of amplitude correlations should be taken into account when interpreting bulk phenomena. These findings reinforce the need to map local CDW amplitude independently of global phase coherence, and highlight the importance of incorporating spatial disorder and the resulting finite coherence length into realistic models of CDW formation and evolution.

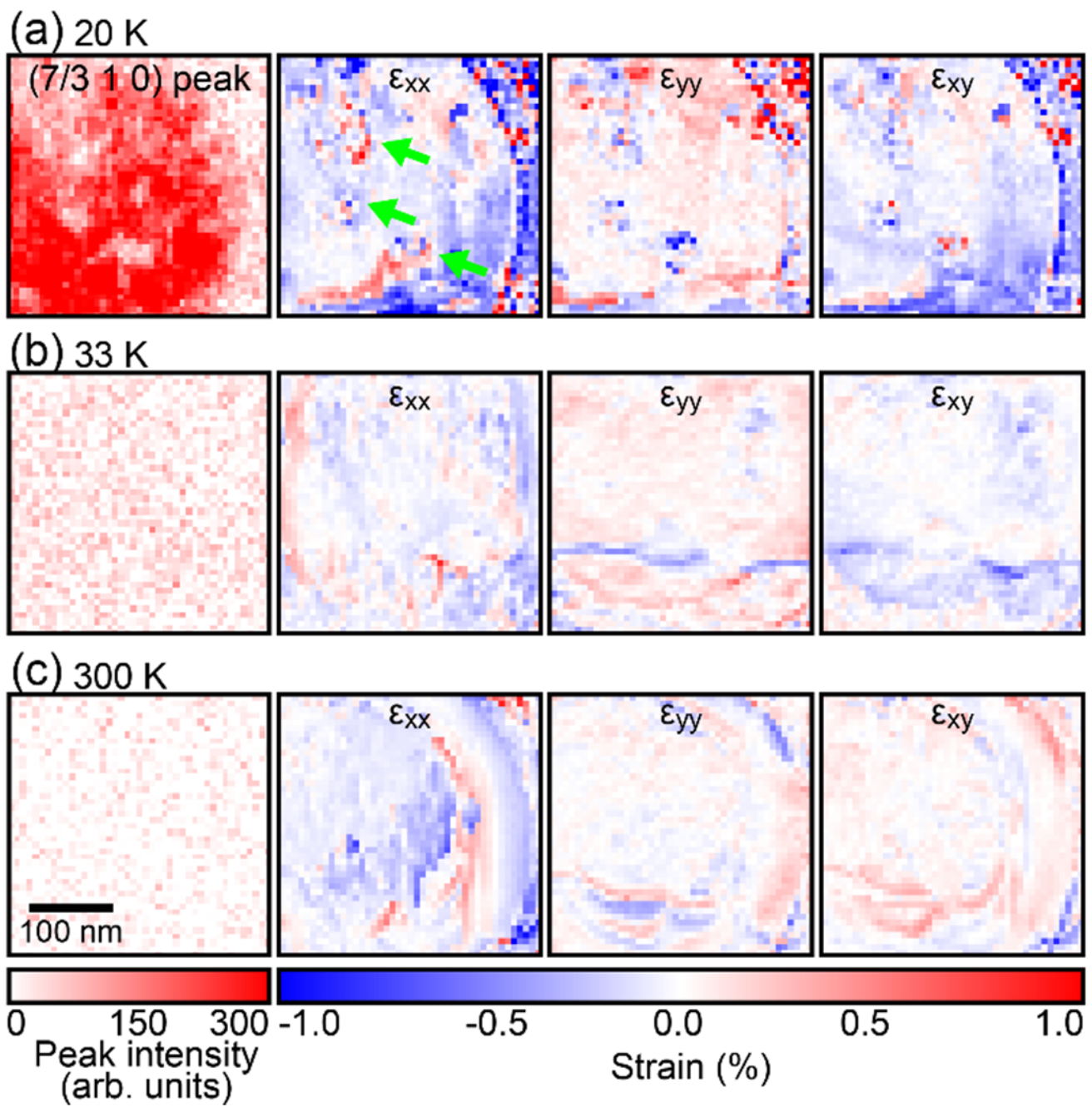

FIG 5. 2D strain maps calculated based on the variation of 4D-STEM diffraction peak positions at (a) 20 K, (b) 33 K and (c) 300 K, alongside the CDW peak intensity maps of the (7/3 1 0) reflection at the corresponding temperatures. The CDW intensity is suppressed in localized strain hotspots (green arrows), while it remains strong in regions of weak strain, demonstrating a clear anticorrelation between CDW order and local strain.

In this work, using liquid helium cryogenic 4D-STEM on a 2H-$NbSe_2$ flake, we directly mapped the local CDW amplitude at nanometer resolution and quantified its spatial correlations. We find that a finite local amplitude is already established at 45 K, well above $T_{CDW}$, and that the associated correlation length grows markedly on cooling, reaching ~110 nm at 20 K. The CDW amplitude is strongly anticorrelated with the local strain field, identifying strain as a local suppressor that limits the extent of amplitude correlations. Our results also demonstrate that amplitude order can persist without global phase coherence (a pseudogap-like regime) and highlight the value of amplitude-specific, spatially resolved probes. More broadly, our approach provides a general pathway to quantify real-space correlation lengths and disorder-limited coherence in complex electronic states. It is readily extendable to fluctuating charge orders, pseudogap phenomena, and emergent textures in unconventional superconductors, correlated oxides, and heavy-fermion materials. Together, these results establish cryogenic 4D-STEM as a powerful platform for probing spatially inhomogeneous quantum order.

## ACKNOWLEDGMENTS

The authors thank E.-G. Moon for helpful discussions. This research was supported by the National Research Foundation of Korea (NRF) Grants (RS-2023-00208179 and RS-2025-02243032) funded by the Korean Government (MSIT). Y.Y. also acknowledges the support from the KAIST singularity professor program. S.B.L. was supported by NRF Grant (2021R1A2C109306013) and Nano Material Technology Development Program through the NRF funded by MSIT (RS-2023-00281839). The 4D-STEM, ADF-STEM and EELS experiments were conducted using a double Cs corrected Titan cubed G2 60-300 (FEI) and Spectra Ultra (ThermoFisher) equipment at KAIST Analysis Center for Research Advancement (KARA). Excellent support by Hyung Bin Bae, Jin-Seok Choi and the staff of KARA is gratefully acknowledged. We declare that the authors utilized the ChatGPT (https://chat.openai.com/chat) for language editing purpose only, and the original manuscript texts were all written by human authors, not by artificial intelligence.

## AUTHOR CONTRIBUTIONS

Y.Y. conceived the idea and directed the study. S.L. and Y.K. synthesized the $NbSe_2$ single crystal. W.C. and H.Y. exfoliated the flake and prepared the TEM specimen. S.H., J.O., and Y.Y. conducted the 4D-STEM experiments. Y.Y. analyzed the experimental data. J.P. and S.B.L. conducted the mean field simulation. C.O. contributed to the interpretation of the experimental results. J.P., S.B.L. and Y.Y. wrote the manuscript. All authors commented on the manuscript. S.H., J.O. and J.P. contributed equally to this work.

## COMPETING INTEREST

S.H., J.O., H.Y., Y.K., S.B.L. and Y.Y. have a patent application (Korea, 10-2025-0059554), which disclose the method for measuring charge density wave orderings in quantum materials. The remaining authors declare no competing interests.

## APPENDIX: DETAILED METHODS

The $NbSe_2$ single crystal was synthesized using chemical vapor transport method. Nb powder (99.999%) and Se powder (99.999%), with a slight excess of Se as the transport agent, were sealed in an evacuated quartz tube and heated at 800°C–900°C for three weeks. The transport measurements for the bulk crystal showed CDW transition temperature of $T_{CDW}$ = 33.5 K and superconducting transition temperature of $T_C$ = 7.4 K (see Supplemental Material for details about the transport characterization and sample quality). The crystal samples were mechanically exfoliated to $Si/SiO_2$ (300 nm) with the scotch tape method and subsequently transferred onto a holey $SiN_x$ grid [Model NH005D03 from Norcada, 200 nm thick $SiN_x$ membrane with a 25 × 25 array of holes approximately 450 nm in diameter; see Fig. 1(b)] with the dry transfer method using the PDMS-PVC (food wrap from Riken Technos) as a stamp.

The grid was loaded onto a Gatan HCHST3008 single-tilt liquid helium-based cooling holder (see Supplemental Material for detailed holder design) and measured using a double Cs-corrected transmission electron microscope (Titan Cubed G2 60-300, FEI). The sample was cooled down to a base temperature of 20 K, and 4D-STEM measurements were conducted consecutively at base temperatures of 20 K, 25 K, 27 K, 29 K, 31 K, 33 K, 35 K, 37 K, 40 K, 45 K, and 300 K. The same grid hole was used for measurements at all temperatures; however, slight variations in the field of view between datasets at different temperatures occurred due to stage drift. Also, the temperatures were recorded at the copper base connecting the helium cooling line to the sample grid (see Supplemental Material for detailed holder design), and the actual sample temperature may differ from these recorded values. The 4D-STEM diffraction patterns were acquired using a K3 Base IS detector (Gatan). The microscope was operated at an acceleration voltage of 300 kV, with a convergence semi-angle of 0.1 mrad. The inner and outer collection angles of the ADF detector were set to 10 mrad and 59 mrad, respectively.

The 4D-STEM experiment scanned 180 × 180 probe positions with a scanning step size of 1.8 nm, and diffraction patterns with k-space dimensions of 511 × 720 pixels were obtained for each probe position, where each k-space pixel corresponds to approximately 17 $\mu m^{-1}$. The beam current was set at 6 pA, and the dwell time for each probe position was 6 ms.

Thermal drift during 4D-STEM acquisition was measured to be ~2.8 Å/s, and the resulting upper-bound errors were estimated and incorporated into Fig. 4(b,d) as error bars. The thermal and cryogen-flow-induced vibration of our liquid helium holder was estimated to be approximately 1 Å from atomic-resolution ADF-STEM imaging at 15 K. While this limits atomic-scale imaging, it does not affect our 4D-STEM diffraction-based correlation analysis, which operates at a spatial resolution of a few nanometers.

The thickness of the exfoliated specimen was determined using the standard EELS log-ratio method [42] with an acceleration voltage of 300 kV, a beam current of 70 pA, a dwell time of 0.2 ms per probe position, and a scan comprising 100 × 100 probe positions with a 5 nm step size.

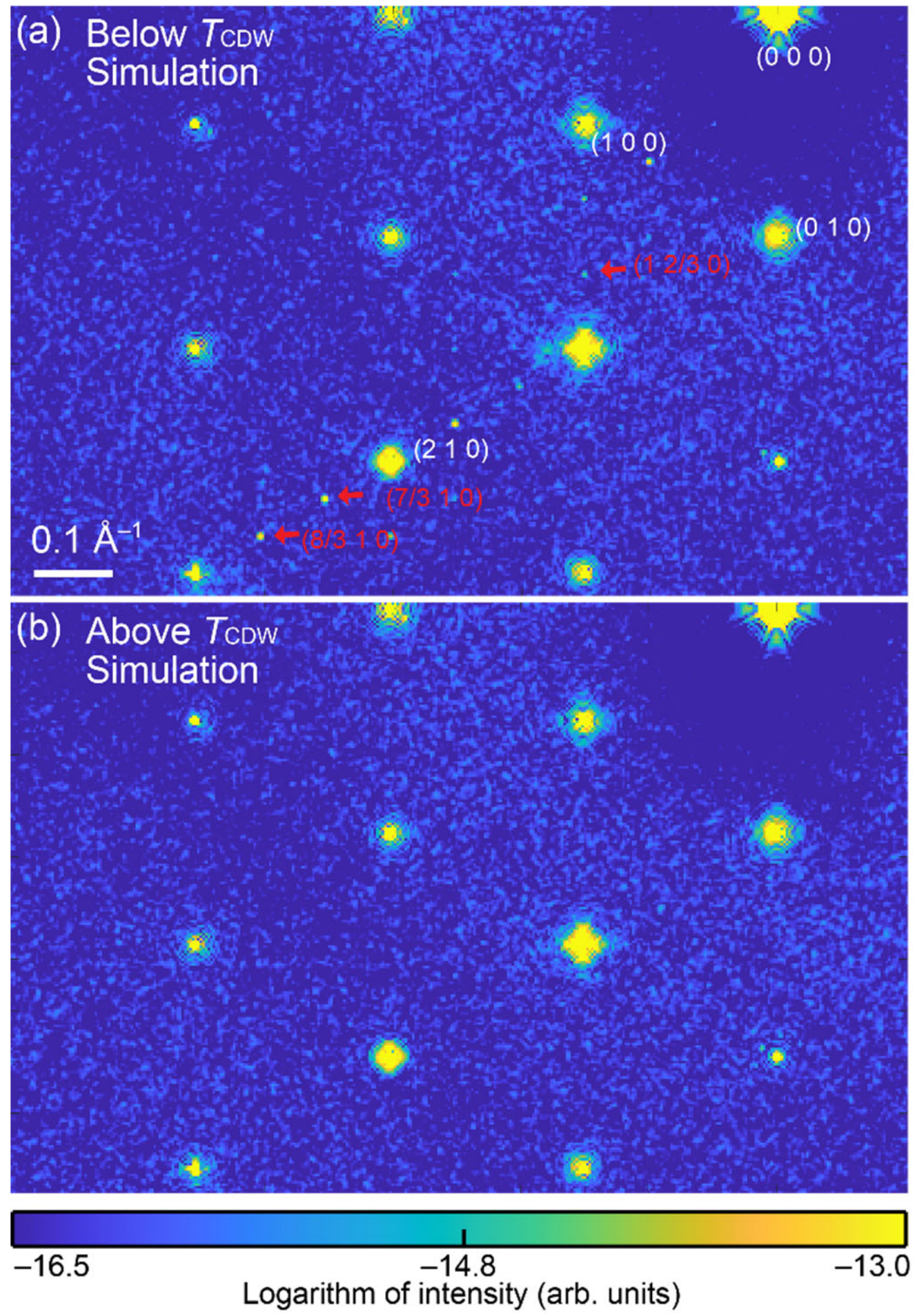

FIG A1. (a) Simulated electron diffraction pattern of 2H-$NbSe_2$ below the CDW transition temperature, obtained using the structural model in [43]. (b) Simulated electron diffraction pattern of 2H-$NbSe_2$ above the CDW transition temperature. Experimental and simulated diffraction patterns are presented with consistent color scales within their respective categories.

The measurement was conducted on a neighboring hole located adjacent to the target hole where the 4D-STEM data were collected.

The effect of electron beam irradiation during 4D-STEM acquisitions was evaluated by separate room temperature TEM measurements, which showed no evidence of beam damage under our experimental conditions (see Supplemental Material for details). The Se vacancy concentration was further estimated from room temperature ADF-STEM measurements combined with thickness determination by EELS, and we obtained a value of 0.030 ± 0.016 % (see Supplemental Material for details).

Electron diffraction pattern simulations were performed using the multislice method implemented in the Prismatic software [44–47]. The simulation parameters included a 300 keV electron energy, a 0.1 mrad convergence semi-angle, a slice thickness of 2 Å, 8 frozen phonon configurations, and an atomic potential sampling of 0.1 Å. Lens aberrations were not considered. The atomic structures below and above $T_{CDW}$ for the simulations were constructed based on previously reported structures [43]. The multislice simulation results depicted in Fig. A1 shows that a ~3° sample tilt along <110> direction results in a diffraction pattern that closely matches the experimentally measured one (Fig. 2).

To improve the signal-to-noise ratio, the 4D-STEM datasets were binned 4 × 4 in real space. As a result, the measured region is represented by 45 × 45 real-space pixels with a pixel size of 7.2 nm. Peak signals were extracted from each diffraction pattern by identifying the locations of (7/3 1 0) and (8/3 1 0) CDW peaks based on strongest integer-index peak positions, and subtracting the local background signal from the CDW peak intensities (see Supplemental Material for more details). Two-dimensional autocorrelation $\langle\varphi(\mathbf{r})\varphi(\mathbf{r}_0+\mathbf{r})\rangle$ was calculated for each CDW intensity map (see Supplemental Material for example of the calculated 2D autocorrelation). The 2D autocorrelation intensities were radially averaged within defined distance intervals using a bin size of ~7 nm. The resulting values were normalized so that the autocorrelation at zero distance equals unity, producing the curves in Fig. 4. For correlation length estimation, the square of the mean value $\langle\varphi(\mathbf{r})\rangle^2$ was first subtracted from the unnormalized autocorrelation curve, followed by normalization at zero distance. The correlation length was then defined as the radial distance where the mean-subtracted autocorrelation decays to $1/e$. The error bars in Fig. 4(a,c) were determined as the standard deviation during the radial averaging.

We quantified strain variations from the 4D-STEM data by first identifying the (1 0 0) and (0 1 0) peak positions of the averaged diffraction pattern (over the full scanned area) to construct a reference 2 × 2 matrix $\mathbf{G}_0$. For each probe position, the corresponding matrix $\mathbf{G}$ was obtained from the peak positions in the diffraction pattern. The deformation gradient was then calculated as $\mathbf{F} = \mathbf{G}_0\,\mathbf{G}^{-1}$, and the 2D strain tensor was given by $\boldsymbol{\varepsilon} = 1/2\,(\mathbf{F} + \mathbf{F}^{T}) - \mathbf{I}$.

## APPENDIX: THEORETICAL SIMULATION

To simulate the physical behavior discussed in the main text, we employed a mean-field theory approach based on the following many-body Hamiltonian $H = t\sum_{<i,j>} c_i^{\dagger} c_j + \sum_{<i,j>} V_{ij} n_i n_j + \sum_i W_i\, n_i$ , where $c_i^{(\dagger)}$ is an electron annihilation (creation) operator at site $i$, with $n_i = c_i^{\dagger} c_i$, and $< i,j >$ denotes nearest-neighbor pairs. The first term describes tunneling between neighboring sites, the second term accounts for nearest-neighbor Coulomb repulsion with $V_{ij} > 0$ and the third term represents on-site disorder, modeled as an Anderson impurity term with $W_i$ uniformly distributed in the range $-0.05 < W_i < 0$. To simplify the interaction term, we applied a mean-field decoupling to the Coulomb repulsion:$\sum_{<i,j>} V_{ij}\, n_i n_j$into $-2V\langle n_{-q_i}\rangle \sum_{i=1}^{3} n_{q_i} + \mathrm{h.c.}$, where $q_i$ denotes the triple-Q wave vectors. Additionally, we included a chemical potential term to fix the filling. To incorporate strain effects, we introduced four locally strained regions by linearly reducing the interaction strength $V_{ij}$ with a Gaussian profile [standard deviation of five lattice constants; see Fig. A2(a)]. The suppressed $V_{ij}$ locally modulates the CDW order parameter landscape. We solved the resulting single-particle Hamiltonian self-consistently across a range of temperatures.

Figure A2(b,c) show the resulting temperature-dependent CDW order parameters below and above $T_{CDW}$, respectively. Each pixel in Fig. A2(b,c) corresponds to a supercell of $3\mathbf{a}_1 \times 3\mathbf{a}_2$ encompassing 3 × 3 lattice sites. A pronounced anticorrelation between the local strain and the CDW order parameter is observed below $T_{CDW}$, but not above $T_{CDW}$. Figure A3(a) depicts the temperature dependence of the spatial correlation of the CDW order parameter. Below $T_{CDW}$, correlations extend over several pixels, whereas above $T_{CDW}$ they decay within less than a pixel, indicating only short-range order. To quantify the correlation between the strain maps and CDW order parameter maps, we calculated their Pearson correlation coefficient (Pearson's r), which measures the relationship between two random variables X and Y. The formula for the coefficient is given by: $r_{xy} = \frac{\sum(X_i-\bar{X})(Y_i-\bar{Y})}{\sqrt{\sum(X_i-\bar{X})^2\sum(Y_i-\bar{Y})^2}}$ where $\bar{X}$ and $\bar{Y}$ represents the means of variables X and Y. A value of +1 indicates a perfect positive linear relationship while –1 indicates a perfect negative linear relationship. As shown in Fig. A3(b,c), both experiment and simulation exhibit a strong anticorrelation between strain and CDW order below $T_{CDW}$, which disappears above $T_{CDW}$.

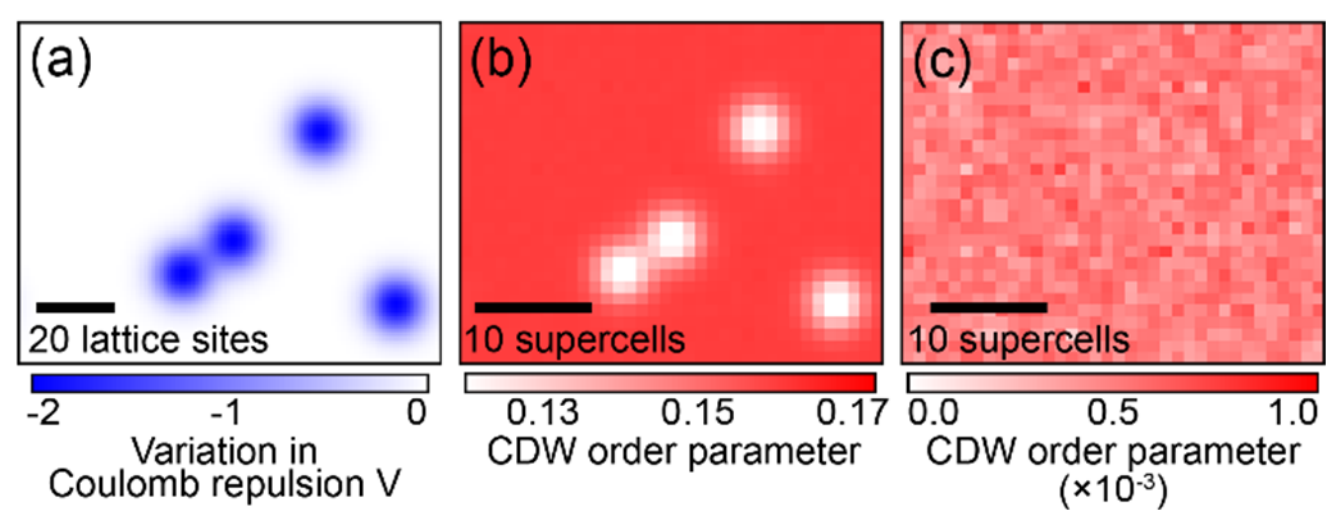

FIG A2. (a) Spatial map of variation in the Coulomb repulsion parameter (effective local strain), modeled with four Gaussian profiles. (b) CDW order parameter at dimensionless inverse temperature $\beta t = t/k_BT = 0.55$, which corresponds to below $T_{CDW}$. (c) Same as (b) but with $\beta t = 0.5$ (above $T_{CDW}$). Regions of high strain in (a) align with suppressed CDW order in (b), indicating an anticorrelation between strain and CDW order. Above $T_{CDW}$, no discernible strain effect is observed.

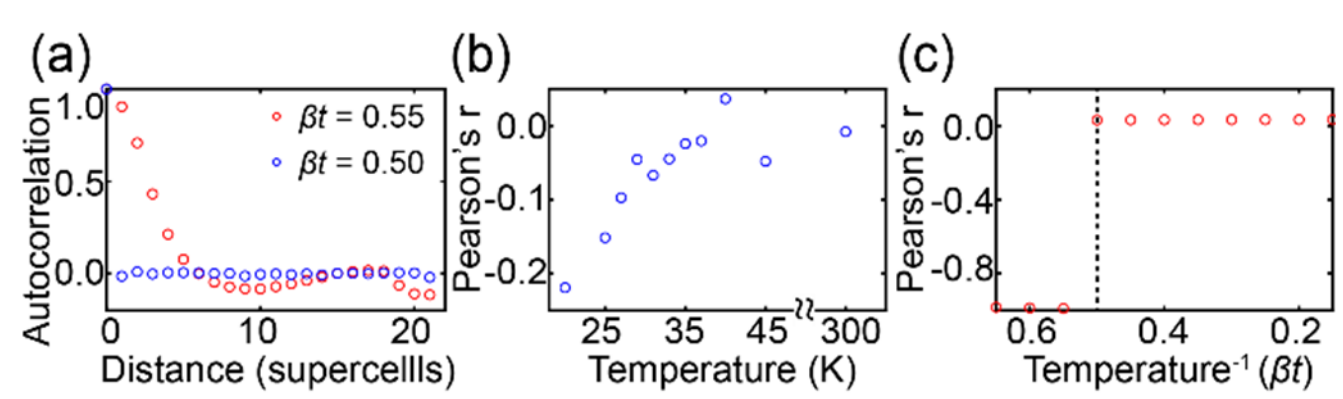

FIG A3. (a) Autocorrelation ($\langle\varphi(\mathbf{r})\varphi(\mathbf{r}_0+\mathbf{r})\rangle - \langle\varphi(\mathbf{r})\rangle^2$) of the simulated CDW order parameters at dimensionless inverse temperatures $\beta t = 0.55$ and 0.5, corresponding to below and above $T_{CDW}$, respectively. (b,c) Pearson's r values obtained from experimental data (b) and simulation (c). Both show a clear negative correlation between strain and CDW order below $T_{CDW}$ (~30 K in experiment, $\beta t$ ~0.5 in simulation).

# Supplemental Material for
# Spatial correlations of charge density wave order across the transition in 2H-$NbSe_2$

Seokjo Hong[1,§], Jaewhan Oh[1,§], Jemin Park[1,§], Woohyun Cho[1], Soyoung Lee[1], Colin Ophus[2], Yeongkwan Kim[1,⁑], Heejun Yang[1,‡], SungBin Lee[1,*] and Yongsoo Yang[1,3,†]

[1]Department of Physics, Korea Advanced Institute of Science and Technology (KAIST), Daejeon 34141, Republic of Korea
[2]Department of Materials Science and Engineering, Stanford University, Stanford, CA 94305, USA
[3]Graduate School of Semiconductor Technology, School of Electrical Engineering, Korea Advanced Institute of Science and Technology (KAIST), Daejeon 34141, Republic of Korea

### Bulk $NbSe_2$ Crystal Characterization

The temperature-dependent resistance of the $NbSe_2$ single crystal was measured using conventional four-probe transport method (Fig. S1). A characteristic hump associated with CDW transition was observed, and the CDW transition temperature was determined to be $T_{CDW}$ = 33.5 K from the temperature derivative of the resistivity near the anomaly [Fig. S1(b)]. The superconducting transition temperature was identified as $T_C$ = 7.4 K from the onset of the resistivity drop. The residual resistivity ratio [ R(300 K) / R($T_{CDW}$) ] was found to be 45.1 ± 4.0, indicating the high quality of the sample.

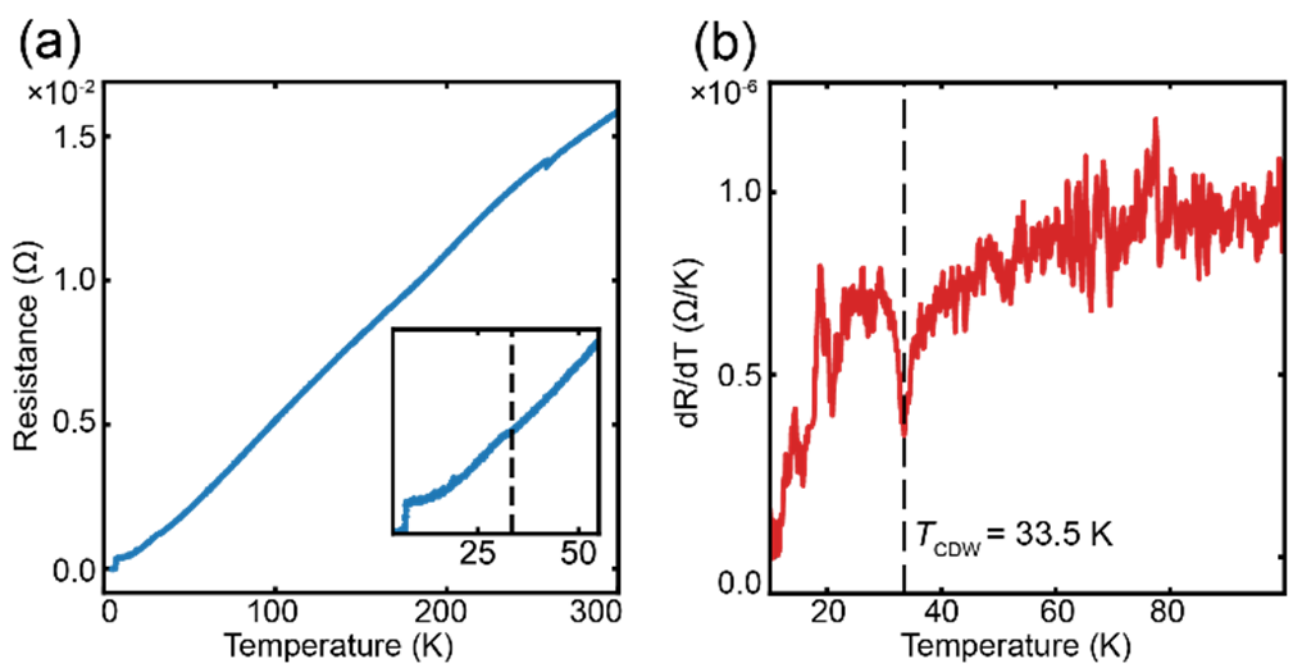

FIG S1. (a) Temperature-dependent resistance of the $NbSe_2$ single crystal. The inset shows a magnified view near the $T_{CDW}$. (b) Temperature derivative of the resistivity curve near the $T_{CDW}$.

### Design of Gatan HCHST3008 Liquid He Cooling Holder

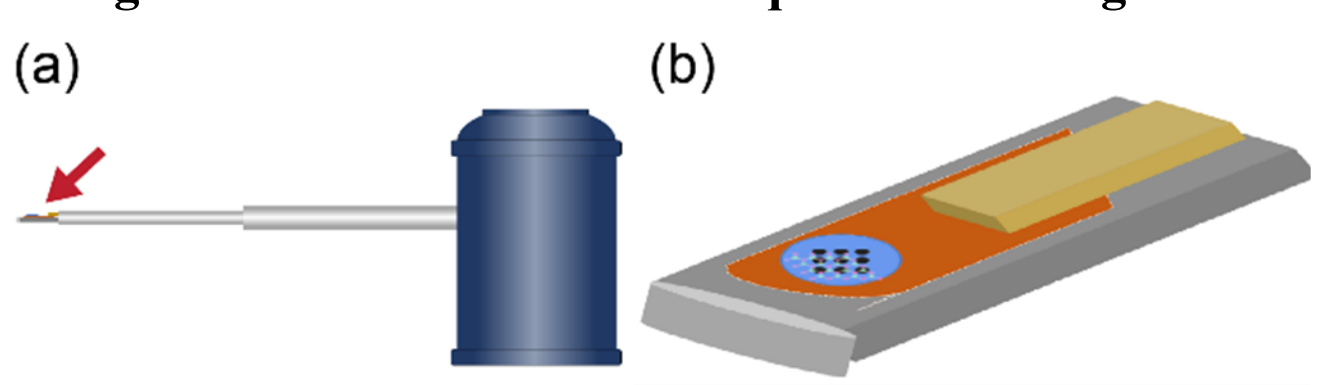

FIG S2. (a) A schematic illustration of the liquid helium-based cooling holder, with a red arrow highlighting the tip region where the specimen is loaded. (b) An enlarged view of the tip region, showing the holey $SiN_x$ sample grid represented as a blue ellipse with black dots. The yellow block indicates the holder base where the temperature was measured.

### Details on CDW Peak Intensity Analysis

Peak signals were extracted from each diffraction pattern as follows: first, the three strongest integer-index peaks were identified in each pattern, and the (1 0 0) and (0 1 0) reciprocal lattice vectors were determined based on their positions. Using these reciprocal lattice vectors, the (7/3 1 0) and (8/3 1 0) CDW peak positions were located. The raw peak signal for each peak was calculated as the average k-space pixel intensity within a distance of 3 pixels from the identified peak position. The background signal for each peak was determined as the average intensity of k-space pixels located between 3 and 5 pixels away from the identified peak position [see Fig. S3(a)]. Finally, the peak intensity was calculated by subtracting the background signal from the raw peak signal.

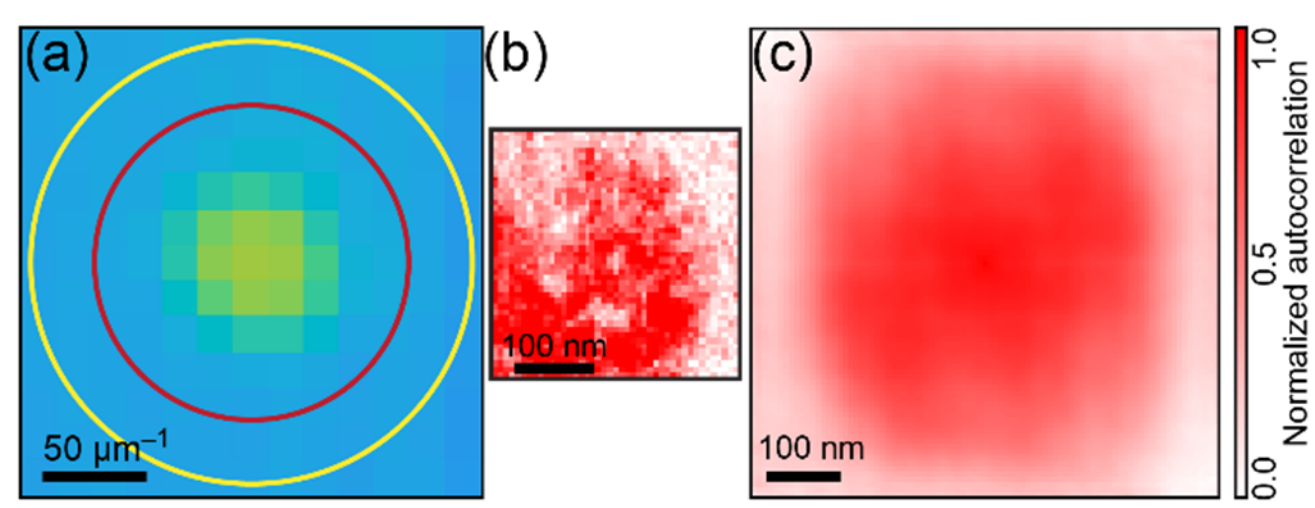

FIG S3. (a) A representative example of a CDW peak observed in the experimentally measured diffraction pattern. The red circle highlights the signal region, while the area between the red and yellow circles defines the background region. (b) An example of the CDW peak intensity map. (c) A 2D autocorrelation derived from the peak intensity map in (b).

### Assessment of Beam-Induced Specimen Damage

The beam condition during each 4D-STEM acquisition corresponds to approximately 700 electrons/$Å^2$, which is well below typical damaging levels for chalcogenide specimens. To confirm that our conditions are safe, we tested ADF-STEM and EELS on a similarly prepared $NbSe_2$ specimen before and after ten consecutive 4D-STEM acquisitions with a slightly higher screen current (10 pA) compared to our cryogenic experiments (6 pA), while keeping all other beam and scan conditions identical. The before and after images and spectra show no sign of beam-induced change (Fig. S4).
During the beam damage assessment, we utilized the Titan double Cs-corrected transmission electron microscope (Titan cubed G2 60-300, FEI) for both ADF-STEM and EELS experiments at an acceleration voltage of 300 kV with a

⁑ Contact author: yeongkwan@kaist.ac.kr ‡ Contact author: h.yang@kaist.ac.kr
* Contact author: sungbin@kaist.ac.kr † Contact author: yongsoo.yang@kaist.ac.kr
§ These authors contributed equally to this work.

convergence semi-angle of 18.0 mrad. Atomic-resolution ADF-STEM images were acquired with a beam current of ~30 pA, a dwell time of 3.0 μs per probe position, and a scan comprising 1024 × 1024 probe positions with a step size of 17.5 pm. The inner and outer collection angles of the ADF detector were set to 38 mrad and 200 mrad, respectively.

EELS data were acquired in dual-EELS mode using a GIF Quantum 965 spectrometer, with a beam current of ~30 pA, a dwell time of 1 ms per probe position, and a 60 × 60 probe scan with a step size of 4.0 nm. The collection semi-angle was 100 mrad. Low-loss and high-loss EELS spectra were collected over the energy ranges of −20.0 eV to 184.7 eV and 45.0 eV to 249.7 eV, respectively, with an energy dispersion of 0.10 eV. To precisely compare any beam induced electronic changes, energy drift correction and background subtraction were applied after data acquisition. Energy drift correction was performed by aligning the zero-loss peak in the low-loss spectra, and background subtraction was then applied to remove the pre-edge background.

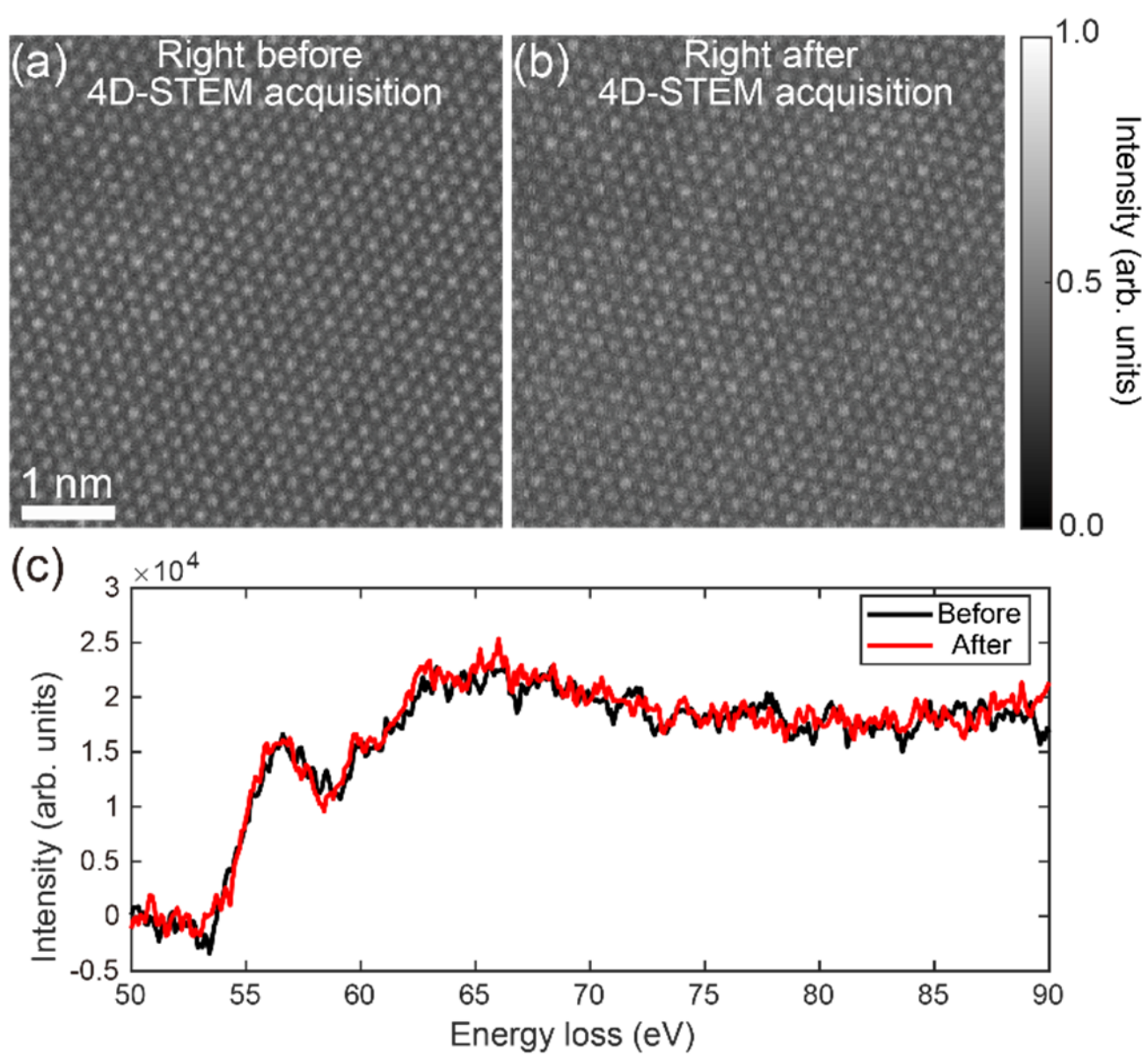

FIG S4. Atomic resolution ADF-STEM images for an $NbSe_2$ flake, acquired (a) before and (b) after ten consecutive 4D-STEM scans. (c) EELS spectra for Se M edge before and after the 4D-STEM scans.

## Estimation of Se Vacancy Concentration

We estimated the Se vacancy concentration using atomic-resolution ADF-STEM and EELS experiments conducted at room temperature using a Spectra Ultra microscope (ThermoFisher), with the $NbSe_2$ single crystal exfoliated and transferred onto a $SiN_x$ grid. The ADF-STEM imaging was performed at an acceleration voltage of 300 kV with a beam current of ~50 pA, a dwell time of 500 ns per probe position, and a scan comprising 4096 × 4096 probe positions with a step size of 3.0 pm. The convergence semi-angle was set to 21.4 mrad, and ADF detector collection angles were between 63 mrad and 200 mrad. The EELS measurement for thickness determination was carried out at 300 kV under conditions of ~100 pA beam current, 0.2 ms dwell time per probe position, and a 38 × 39 probe scan with a step size of 3.0 nm. The convergence semi-angle was 21.4 mrad.

To quantitatively analyze the positions and intensities of atomic columns from the ADF-STEM image, the image was first processed using a Gaussian filter with standard deviation of 30 pm to suppress noise [see Fig. S5(a)]. Local intensity maxima pixels were then identified, while enforcing a minimum distance constraint of ~1.7 Å (around half of the Se-Se distance) between the identified local maxima, resulting in 3,754 candidate atomic column positions. For each candidate site, 2D Gaussian fitting was applied twice: an initial fitting on a 63 × 63 pixel box (~1.9 Å × 1.9 Å) around the candidate site, and a refined fitting on 23 × 23 pixel box (~0.7 Å × 0.7 Å) around the initially fitted peak position. The raw-image (before Gaussian smoothing) intensities within a 29 × 29 pixel box (~0.9 Å × 0.9 Å) centered on the rounded coordinates of each fitted columns positions were averaged to represent the intensity of each column. Each atomic column was classified as Nb and Se columns, respectively, based on the known $NbSe_2$ crystal structure. Through this process, we identified 1,251 Nb and 2,503 Se atomic columns, as shown in Fig. S5(b).

To estimate the Se vacancy concentration, we further plotted the histogram of the averaged Se atomic column intensities. For a perfect crystal without defects, the histogram should be symmetric and Gaussian-like, reflecting thermal and Poisson noise effects. In Fig. S5(c), however, the histogram is not fully symmetric and exhibits a longer tail with higher counts in the low-intensity region. This indicates the presence of columns containing Se vacancies, which contribute to the excess counts in the low-intensity bins. By fitting a Gaussian to the right half of the histogram and subtracting this fitted function from the left half, we estimated the expected number of vacancies per bin. Using the EELS-based thickness measurement (17.0 ± 0.9 nm) and assuming the presence of two atomic layers per unit cell, we determined a total of 25 layers across the sample thickness. Each layer was assumed to contain one Se atom per column on average. This enabled the calculation of the total number of Se atoms within the field of view and allowed us to estimate the Se vacancy concentration to be 0.030% ± 0.016%.

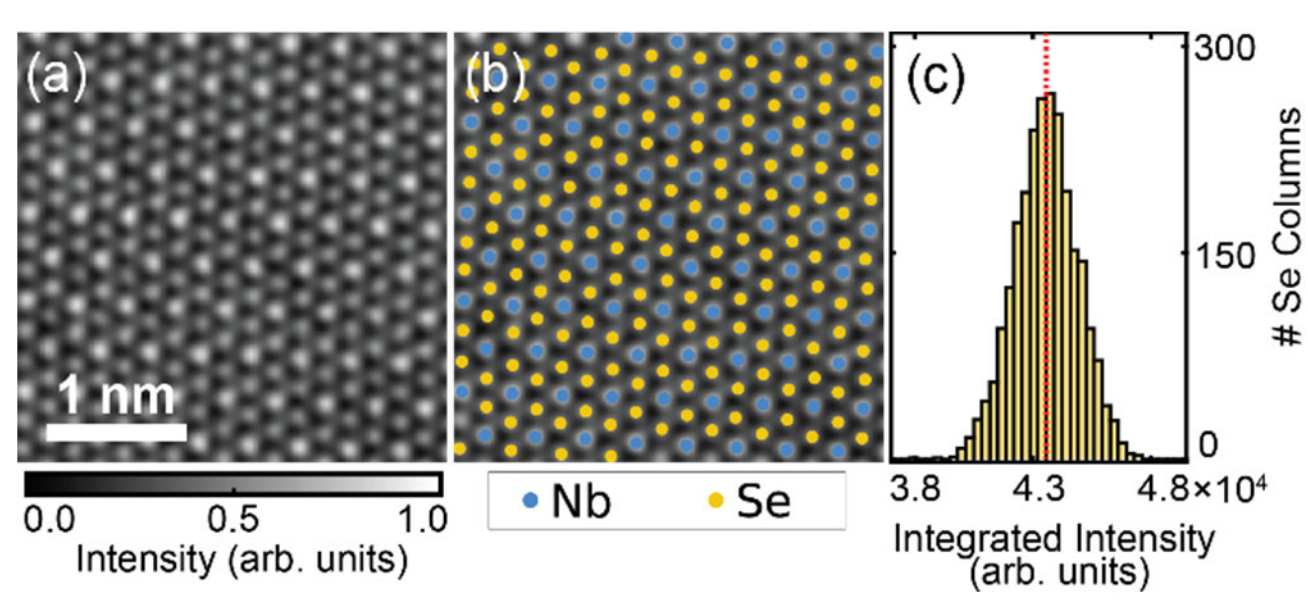

FIG S5. (a) An atomic-resolution ADF-STEM image of $NbSe_2$ flake after Gaussian smoothing. (b) Same image as (a), with the identified Nb and Se column positions overlaid. (c) Intensity histogram of Se atomic columns.

**A Note on the Position of (0 0 0) Peak and Beam Stopper**

During 4D-STEM acquisition, the diffraction was shifted to include the brightest (7/3 1 0) and (8/3 1 0) CDW peaks on the detector. Consequently, the (0 0 0) peak was outside the field of view, and the beam stopper was placed on the next-brightest (1 1 0) peak to maximize the detector dynamic range.